# How Turbulent is the Magnetically Closed Corona?

James A. Klimchuk and Spiro K. Antiochos

Heliophysics Science Division
NASA Goddard Space Flight Center
Greenbelt, MD  20771  USA

## Abstract

We argue that the magnetically closed corona evolves primarily quasi-statically, punctuated by many localized bursts of activity associated with magnetic reconnection at a myriad of small current sheets. The sheets form by various processes that do not involve a traditional turbulent cascade whereby energy flows losslessly through a continuum of spatial scales starting from the large scale of the photospheric driving. If such an inertial range is a defining characteristic of turbulence, then the magnetically closed corona is not a turbulent system. It nonetheless has a complex structure that bears no direct relationship to the pattern of driving.

## 1.  Introduction

The surface of the Sun – the photosphere – undergoes incessant chaotic motions associated with the convective transport of energy from below. A portion of that energy is transported into the corona by the magnetic field and heats the gas to million-degree temperatures. Some of the surface flows change direction rapidly and launch Alfven waves that subsequently dissipate. Most are much longer-lived, however, and provide heating in a different manner. The details of how this occurs are the subject of this "perspective" article. We deal specifically with the magnetically closed corona, where field lines are rooted to the surface at both ends, allowing magnetic stresses to develop readily. This is the realm of active regions and the quiet Sun. Magnetically open coronal holes and the solar wind cannot support significant stress and are heated primarily by waves (Cranmer et al., 2017). Although still debated, it seems that non-wave heating dominates in the closed corona, especially within active regions; see the recent reviews of observations and models by Van Doorsselaere et al. (2020) and Viall et al. (2021) as well as the recent work by Howson et al. (2020).

There two competing views for how magnetic stresses are built up and released in the corona. In one picture, the evolution is mostly quasi-static but punctuated by many small, localized bursts of activity. The other picture is fully dynamic, with flows having a wide range of spatial scales completely filling the system. This is the turbulence view of the corona. Electric current sheets play an important role in both cases. They are the sites of magnetic reconnection events – often called nanoflares (Parker, 1983) – that are the primary agents of heating. The fundamental difference between the two pictures is how current sheets form. Both possibilities may occur in the actual corona, but our perspective is that one of them dominates.

In driven hydrodynamic systems with small viscosity, nonlinear interactions cause large-scale flows to break up into smaller and smaller eddies. The system organizes such that kinetic energy flows without loss from the largest scales at which it is injected, through a continuum of ever-smaller scales, ultimately reaching a scale where gradients are steep enough that viscous heating is effective. The intermediate range of lossless energy cascade is known as the inertial range and is a fundamental property of turbulence (Kolmogorov, 1941; Biskamp, 1993). In MHD systems, turbulence is more complicated due to the presence of a magnetic field, but an inertial range is still a defining feature. Magnetic reconnection is an additional energy-dissipation mechanism, obviously not present in hydrodynamics. Reconnection occurs at current-sheet scales larger than the viscous and resistive dissipation scales, but still much smaller than the scale of the driving. Spatially and temporally intermittent behavior is present in hydrodynamic turbulence, but it is much more prominent in MHD turbulence because of reconnection.

Photospheric driving can create current sheets by at least three processes different from a turbulent cascade:

First, even smooth large-scale flows immediately create current sheets at magnetic topological boundaries called separatrix surfaces and quasi-separatrix layers (QSLs) (Priest et al., 2002). These boundaries occur because the field is highly fragmented in the photosphere, with much of the flux concentrated in structures known as kilogauss flux tubes. The tubes expand rapidly with height to become space filling in the low-$\beta$ corona; thus, they are separated in the photosphere but in contact in the corona. Such contact boundaries are QSLs. Magnetogram observations indicate that a single active region contains roughly $10^5$ photospheric concentrations of kiloGauss field, implying a comparable number of current sheets in the corona above (Klimchuk 2015). Partial reconnection of adjacent flux tubes will double the number of topologically distinct structures as described in Klimchuk (2015); therefore, we can expect the actual number of sheets in an active region to far exceed $10^5$.

Second, even without the clumping of field described above, a complex evolution for the photospheric flows – including flows that are smooth and large-scale – will cause current structures to form in the corona and thin at an exponential rate (e.g., van Ballegooijen, 1986; Pontin and Hornig, 2020). A stagnation point flow is representative of the basic effect. The footpoints of two nearby field lines may move together for some time, but then diverge as the stagnation point is approached. A quantitative demonstration of this effect can be found in Antiochos and Dahlburg (1997).

Third, current sheets can form highly dynamically from coronal instabilities. The resistive internal kink instability is one example, whereby the distributed volume currents in a twisted flux tube are rapidly converted into multiple thin sheets (Hood et al., 2009). Also, when reconnection is patchy, multiple new current sheets are dynamically created through a process known as reconnection driven current filamentation (Karpen et al., 1996).

We conclude that current sheets and the resulting coronal heating can readily occur without the need of a turbulent cascade; but how can we determine which of these two competing views of the corona is correct: fundamentally turbulent or mostly quasi-static? Important insight may be provided by the spatial energy spectrum, as we now discuss.

## 2. Energy Spectra

Turbulence theory predicts that the lossless energy cascade at intermediate scales – the inertial range – has a distinctive spectrum of the form: $E(k) \propto k^{-\alpha}$, where $E$ is energy and $k$ is wavenumber. The spectral index $\alpha$ has the famous value 5/3 for the kinetic energy in hydrodynamic turbulence (Kolmogorov, 1941). For MHD turbulence in open systems, there is equipartition between magnetic and kinetic energies, and they both have the same spectral index, ranging between 1.5 and 2 depending on whether the turbulence is weak or strong, isotropic or anisotropic. We must keep in mind that the corona is a closed system and may have different spectral properties from an open system if turbulence exists (Rappazzo et al., 2007, 2008; Rappazzo & Velli 2011).

Many numerical simulations have been performed to study the coronal energy spectrum resulting from slow photospheric driving (Hendrix and Van Hoven, 1996; Dmitruk et al., 2003; Rappazzo et al., 2007, 2008, 2010, 2013; Rappazzo and Velli, 2011). They begin with a uniform magnetic field in a box that spans two opposing boundaries representing positive and negative polarity parts of the photosphere. Essentially, the curved coronal field is straightened out. Flows are imposed at the boundaries that have a random aspect representative of photospheric convection. To avoid the generation of waves, the flows are either steady or their correlation times are long compared to the Alfven travel time.[1]

In all cases, the system responds in a well-behaved manner initially. Magnetic stresses slowly increase as the field becomes twisted and tangled, but the evolution remains everywhere quasi-static even when the stresses become large, thereby demonstrating that nonlinear interactions do not break up smooth flows into smaller and smaller eddies, as occurs with hydrodynamic turbulence. The stiffness of the line-tied, low-$\beta$ magnetic field lines resists any bending by the plasma. Significant dynamics can result only from imbalances between the two magnetic forces - tension and magnetic pressure gradient - not from fluid effects.

The system becomes dynamic only when a magnetic instability sets in. Kinking and tearing instabilities are the most common. The critical level of twist for kinking depends on the twist profile (Bareford 2010); for example, our simulations require three full turns for instability (Klimchuk et al., 2010).[2] The length-to-diameter aspect ratio in our model is 8, so the field is highly stressed and nonlinear with $B_\phi/B_z \sim 1$.

When the initial instability occurs, it usually produces magnetic complexity and current sheet formation. This leads to reconnection, which produces additional complexity leading to more reconnection and a proliferation of current sheets. Eventually, the system settles into a statistical steady state in which the energy released by the reconnection events balances the Poynting flux input at the boundaries from the work done on the field by the driving.

Kinetic and magnetic energy spectra for this system are computed from the square of the Fourier transforms of the transverse (i.e., perpendicular to the initial magnetic field direction) components of velocity and magnetic field. They are sometimes called the fluctuating components, though this can be misleading, as we argue below. The spectra from multiple studies are found to obey approximate power laws at intermediate wavenumbers, with the kinetic index being much smaller than the magnetic index: $\alpha_K \approx 0.5$ and $\alpha_M \approx 2$-3. Furthermore, magnetic energy exceeds kinetic energy by roughly two orders of magnitude: $E_M \approx 100\ E_K$ (Hendrix and Van Hoven, 1996; Dmitruk et al., 2003; Rappazzo et al., 2007, 2008, 2010, 2013; Rappazzo and Velli, 2011). These results differ dramatically from the theoretical predictions for an open turbulent system. They are nonetheless cited as evidence for an inertial range, with the discrepancy being attributed to the effects of line-tying. We claim that this interpretation is premature and propose an alternative explanation below.

---

[1] Some studies intentionally launch waves (e.g., van Ballegooijen, et al., 2011; Howson, et al. 2020), but those are not considered here.

[2] We use full MHD and impose a driving pattern in which vorticity is non-constant on streamlines – a condition said to be necessary to activate the nonlinear terms in the equations (Rappazzo et al. 2008). Despite this, no "nonlinear dynamics" occurs until kinking sets in.

Note that observationally derived magnetic and kinetic energy spectra do not exist for the magnetically closed corona like they do for the solar wind. Remote sensing measurements of the magnetic and velocity vectors are neither sufficiently accurate nor on sufficiently small scales. Furthermore, the optically thin nature of the corona means there is spatially averaging along the line of sight.

The Fourier power spectrum (square of the transform) is usually interpreted as a measure of the distribution of the different spatial scales that are present in a system. This is not necessarily the case, however. Spatial discontinuities have a power spectrum that is a power law with spectral index 2 (Nahin 2001). This is demonstrated by the two simple examples of Figure 1. On the top left is a step function that might represent the abrupt jump in the transverse component of magnetic field across a current sheet. The corresponding power spectrum is below. The right side shows the magnetic profile and spectrum for 993 sheets with random spacing and random sign (positive or negative jump in the field). Both spectra are nearly straight lines in the log-log plots, but deviate slightly at high wavenumbers, likely due to the discretization in the model. Linear fits have slopes of -1.8 if the full range is included, and very close to -2.0 if high wavenumbers are excluded (above $10^2$ on the left and above $10^3$ on the right).

The fact that discontinuities have spectral index of 2 has led some to suggest that the observed power spectra of the solar wind may be an indication of discontinuities (current sheets) rather than a turbulent cascade (Roberts and Goldstein, 1987; Borovsky, 2010). We propose that the MHD simulations representing the magnetically closed corona, discussed above, can be interpreted in the same way. This would explain why the spectral indices for magnetic and kinetic energy differ greatly from each other and from the expected value. It would also explain why the magnetic energy is two orders of magnitude larger than the kinetic energy, rather than equal as expected for turbulence.

The background field in the simulations is initially uniform and potential, but a large transverse component develops as stresses are built up quasi-statically during the pre-dynamics phase. Presumably, some of these stresses remain when the system transitions to a statistical steady state, especially if the driving is steady, as is often the case. If the transverse component of the background field greatly exceeds that of the dynamic fluctuations, the magnetic spectral index will be primarily a measure of the slowly evolving background rather than the fluctuations. If current sheets are prevalent, the index should be near 2.

This alone does not argue against turbulence. There could be a hidden turbulent part of the magnetic energy that does not affect the index and that is in equipartition with kinetic energy. The problem is that the kinetic index in the simulations is near 0.5, far smaller than the expected theoretical value of 1.5-2. This would need to be rigorously explained, though see the discussion in Rappazzo and Velli (2011). Another possibility, of course, is that there is no hidden turbulent component in the magnetic spectrum, and the velocity spectrum is due to something entirely different from turbulence.

## 3. Discussion

We have described are two competing pictures of the corona in which slow photospheric driving creates a myriad of current sheets that reconnect sporadically to heat the plasma. In the turbulence picture, the corona is a fully dynamic system. Current sheets are produced at the end of a systematic and lossless cascade of energy through a continuum of spatial scales that starts at the large scale of the driving. The competing picture is primarily quasi-static, with current sheets forming by a variety of processes that do not involve an organized system of spatial scales. Many different scales are present, but they are not connected in the manner of an inertial range. In both pictures, the corona has a complex structure that bears no direction relationship to the pattern of driving.

Magnetic and kinetic energy spectra in numerical simulations have been offered as evidence for the turbulence picture. In our opinion, however, the properties of these spectra argue as much against turbulence as for it. Furthermore, the well-behaved nature of the simulations before the first occurrence of a magnetic instability – often far into the nonlinear regime – is difficult to reconcile with traditional turbulence.

Some may argue that the quasi-static picture is just another type of turbulence because of its complex nature. We recommend against this label and suggest that the term be reserved for systems in which a majority the volume is undergoing a lossless energy cascade beginning at the scale of the driving. Magnetic reconnection can be very chaotic depending on the properties of the current sheet (Daughton et al., 2011; Huang and Battacharjee, 2016; Leake et al., 2020), and it is possible that the quasi-static picture we advocate includes localized regions where an inertial range is temporarily established. Whether these regions account for a sizable fraction of the coronal volume depends on several factors: the number density of current sheets; the frequency with which they reconnect; the fraction of events that develop an inertial range; and the rate at which the activity decays. Note that the inertial range would begin at the scale of the current sheet, which is far smaller than the scale of the driving, so this would not fit our definition of a turbulent system in any case.

There is much more work to be done before a comprehensive understanding of the corona is achieved. We believe the quasi-static picture serves as a good foundation. The magnetic spectral index found in simulations can likely be explained, at least in part, by the preponderance of current sheets. Why the index is sometimes significantly larger than 2 is possibly related to the distribution of current sheet spacings, but that has yet to be investigated. The kinetic spectral index of ~0.5 must also be explained. An avalanche-like behavior of reconnection events is one possibility. We must remember that, because of topological complexities such as separators and QSLs, the velocity pattern in the photosphere - which may be dominated by one scale - is translated into a different velocity pattern in corona, likely involving a range of scales. This will affect the shape of the spectrum. Finally, we note that $\alpha_K < 1$ indicates that small spatial scales contain more energy than large scales, which is fundamentally at odds with turbulence. It would be expected, on the other hand, if the dominant flows are due to reconnection rather than being a direct result of the driving.

We thank Marco Velli, Franco Rappazzo, Daniel Gomez, Ben Chandran, Russell Dahlburg, Peter Cargill, Francesco Malara, Andrew Hillier, Aaron Roberts, Giuseppina Nigro, and Vadim Uritsky for a number of helpful discussions in recent years. We also thank the referees for their comments and suggestions.

# References

Antiochos, S. K., & Dahlburg, R. B. 1997, Solar Phys., 174, 5

Bareford, M. R., Browning, P. K., & Van der Linden, R. A. M. 2010, AA, 521, A70

Biskamp, D. 1993, Nonlinear Magnetohydrodynamics (Cambridge; Cambridge Univ. Press)

Borovsky, J. E. 2010, PRL, 105, 111102

Cranmer, S. R., Gibson, S. E., & Riley, P. 2017, Sp. Sci. Rev., 212, 1345

Daughton, W. et al. 2011, Nature Physics, 7, 539

Dmitruk, P., Gomez, D. O., & Matthaeus, W. H. 2003, PhPl, 10, 3584

Hendrix, D. L., & Van Hoven, G. 1996, ApJ, 467, 887

Hood, A. W., Browning, P. K., & Van der Linden, R. A. M. 2009, A&A, 506, 913

Howson, T. A., De Moortel, I., & Fyfe, L. E. 2020, A&A, 643, A85

Huang, Y.-M., & Bhattacharjee, A. 2016, ApJ, 818, 20

Karpen, J. T., Antiochos, S. K., & DeVore, C. R. 1996, ApJL, 460, L73

Klimchuk, J. A. 2015, Phil. Trans. R. Soc. A, 373, 20140256

Klimchuk, J. A., Nigro, G., Dahlburg, R. B., & Antiochos, S. K. 2010, BAAS, 41, 847

Knizhnik, K. J., Uritsky, V. M., Klimchuk, J. A., & DeVore, C. R. 2018, ApJ, 853, 82

Kolmogorov, A. N. 1941, Dokl. Akad. Nauk. SSSR, 30, 299

Leake, J. E., Daldorff, L. K. S., & Klimchuk, J. A. 2020, ApJ, 891, 62

Nahin, P. J. 2001, The Science of Radio (New York; Springer-Verlag)

Parker, E. N. 1983, ApJ, 264, 642

Pontin, D. I., & Hornig, G. 2020, Living Rev. Sol. Phys., 17, 5

Priest, E. R., Heyvaerts, J. F., & Title, A. M. 2002, ApJ, 576, 533

Rappazzo, A. F., & Velli, M. 2011, Phys Rev E, 83, 065401

Rappazzo, A. F., Velli, M., Einaudi, G., & Dahlburg, R. B. 2007, ApJL, 657, L47

Rappazzo, A. F., Velli, M., Einaudi, G., & Dahlburg, R. B.  2008, ApJ, 677, 1348

Rappazzo, A. F., Velli, M., & Einaudi, G.  2010, ApJ, 722, 65

Rappazzo, A. F., Velli, M., & Einaudi, G.  2013, ApJ, 771, 76

Roberts, D. A., & Goldstein, J. L.  1987, JGR, 92, 10105

van Ballegooijen, A. A.  1986, ApJ, 311, 1001

van Ballegooijen, A. A., Asgari-Targhi, M., Cranmer, S. R., & DeLuca, E. E.  2011, ApJ, 736, 3

Van Doorsselaere, T., Srivastava, A. K., Antolin, P. et al.  2020, SSR, 216, 140

Viall, N. M., De Moortel, I., Downs, C., Klimchuk, J. A., Parenti, S., & Reale, F.  2021, in Space Physics and Aeronomy Collection Vol. 1:  At the Doorstep of Our Star:  Solar Physics and Solar Wind, Geophys. Monograph 258, eds: N. E. Raoufi & A. Vourlidas) (AGU; Wiley)

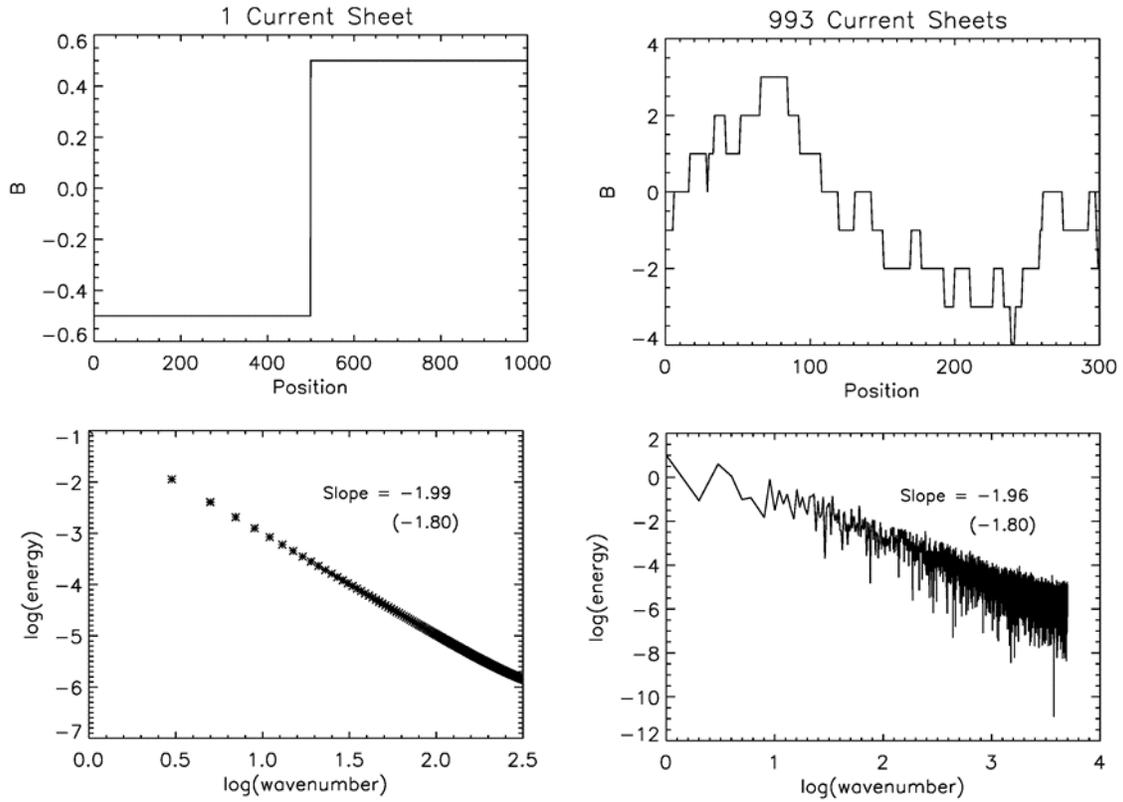

Figure 1: Spatial profile of transverse magnetic field (top) and Fourier power spectrum (bottom) for a single current sheet (left) and 993 currents sheets with random spacing and random sign (right). Only a portion of the profile is shown for the multi-sheet case.